\documentclass[prl,twocolumn,floatfix,superscriptaddress]{revtex4-1}
\usepackage{dcolumn,amsmath}
\usepackage{graphicx}
\usepackage{bm}
\usepackage{hyperref}

\newcommand{\ecm}{\ensuremath{e {\cdotp} {\rm cm}}}

\begin{document}
\title{
Theoretical study of the parity and time reversal violating interaction in solids}

\author{L.V.\ Skripnikov}\email{leonidos239@gmail.com}
\author{A.V.\ Titov}
\homepage{http://www.qchem.pnpi.spb.ru}
\affiliation{Federal state budgetary institute ``Petersburg Nuclear Physics Institute'', Gatchina, Leningrad district 188300, Russia}
\affiliation{Dept.\ of Physics, Saint Petersburg State University, Saint Petersburg, Petrodvoretz 198904, Russia}
\date{\today}

\begin{abstract}
A new theoretical approach to study the properties in solids, which are sensitive to a change of densities of the valence electrons in atomic cores (hyperfine structure constants, parameters of space parity (P) and time reversal (T) violation interaction, etc.) is proposed and implemented. 
It uses the two-step concept of calculation of such properties which was implemented earlier for the case of molecules [Progr.\ Theor.\ Chem.\ Phys.~B 15, 253 (2006)].  
The approach is applied to evaluate the parameter $X$ describing electronic density gradient on the Pb nucleus that is required to interpret the proposed experiment on PbTiO$_3$ crystal [PRA, 72, 034501 (2005)] to search for the Schiff moment of the $^{207}$Pb nucleus because of its high sensitivity to the corresponding P,T-violating interactions. For comparison the $X$ parameter has also been calculated on the Pb nucleus for the $^1\Sigma^+$ state of $^{207}$PbO molecule using the same density functionals as those utilized in PbTiO$_3$ studies. The relativistic coupled-clusters approach with single, double and perturbative triple cluster amplitudes, applicable to a few atom systems and providing high accuracy for $X$, is also applied to the PbO case to estimate the accuracy of density functional studies.
%
\end{abstract}

\maketitle

\section*{Introduction}

The recent identification of the new particle
discovered at the LHC as a Higgs boson with a mass of 125 GeV/c$^2$ completes the picture of particles and forces described by the Standard model \cite{Higgs_boson_birth:13a}. However, it does not mark the end of the story as, unfortunately, the Standard model is an incomplete description of nature. Puzzles still remain, for example, in explaining the existence of dark matter and the matter$-$antimatter asymmetry. Search for the effects of fundamental CP symmetry violation (C is the charge conjugation symmetry and P is space parity) can shed light on the latter problem. Via the CPT theorem (T is the time-reversal invariance), CP violation means also T symmetry nonconservation. In this connection, search for the T,P-parity nonconservation effects \cite{Ginges:04} including the permanent electric dipole moments (EDM) of elementary particles and nuclear Schiff momenta (NSM), becomes now one of the most intriguing problems of modern physics.

Almost half a century ago Sandars \cite{Sandars:65, Sandars:67} and Shapiro \cite{Shapiro:1968} realized that very perspective experiments towards the search of violation of fundamental symmetries can be performed on atoms, molecules and solids containing heavy elements. Though the possible CP violation mechanism within the Standard model generate too small effects, many extensions to the Standard model give rise to such T,P-odd effects which are already in the reach of modern experiments. Recently a molecular experiment on the YbF beam has succeed to obtain a very rigid upper bound on the electron EDM~\cite{Hudson:11a}, $1.05\cdotp10^{-27}\ecm$. Up to now, the best limitation on the nuclear T,P-odd interactions was archived in atomic Hg experiments \cite{Griffith:09}.

The difficulties of all these searches are not only experimental. To interpret the measured data in terms of the electron EDM, nuclear Schiff moments etc.\ one should know a number of parameters, which are determined by the electronic structures of systems under consideration.
These parameters can not be measured, their evaluation constitute a problem of ab initio electronic structure study that is especially challenging for solids.
%
There are other properties of interest in different applications which, similar to the T- and P-odd effects, are sensitive to a change of densities of the \textit{valence} electrons in \textit{atomic cores}: hyperfine magnetic dipole and electric quadruple constants, chemical shifts of X-ray emission lines, volume isotope and M{\"o}ssbauer shifts, etc.\ \cite{Titov:06amin,Lomachuk:13a}.

In the present paper we introduce and implement a new method of calculation of the above mentioned characteristics (describing the state of atoms in solids rather than the chemical bonding
and called below
the ``core characteristics'', ``core properties'' or ``core parameters'' for simplicity) which take account of both relativistic and correlation effects explicitly and, from other side, is valid for the periodic structures. As a first application of this method we have evaluated the parameter $X$ (see below) that is required to study the T,P-violating interaction in the PbTiO$_3$ crystal.

The use of the PbTiO$_3$ crystal to search for the  Schiff moment of the $^{207}$Pb nucleus has recently been suggested by Mukhamedjanov and Sushkov in Ref.~\cite{Mukhamedjanov:05}. According to the authors of ~\cite{Mukhamedjanov:05} one can reach a sensitivity up to
ten orders of magnitude better than the current result for Hg \cite{Griffith:09}.
At 763~K the PbTiO$_3$ crystal undergoes ferroelectric phase transition from cubic 
to tetragonal symmetry.
In the ferroelectric phase, the Pb and Ti atoms are displaced along the tetragonal axis (``c'' on fig.~1) with respect to their non-ferroelectric positions.
In contrast to a similar ferroelectric phase case in BaTiO$_3$, the displacements of the atoms in PbTiO$_3$ are much bigger \cite{Nelmes:85}. This leads to a strong ferroelectricity of PbTiO$_3$ which can induce strong internal effective fields.

\begin{figure}
\label{PTO}
\includegraphics[scale=0.3]{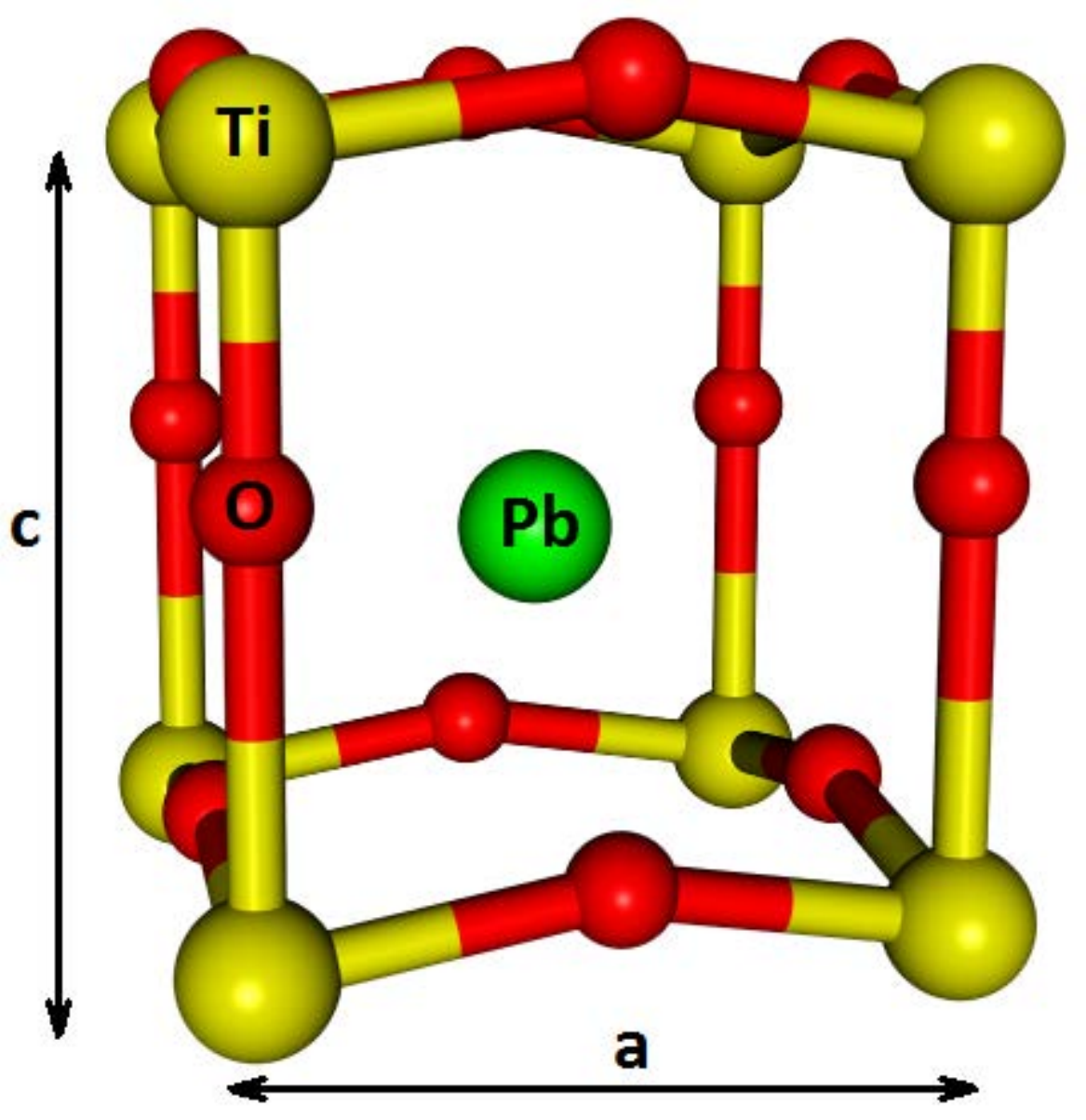}
\caption{(Color online) Structure of PbTiO$_3$ crystal in the ferroelectric phase.
}
\end{figure}

The effective T,P-odd interaction with a nucleus having a Schiff moment can be written in the form \cite{Hinds:80a}
\begin{equation}
     H_{\rm eff}= -6SX \vec{\sigma}_N \cdot \vec{\lambda}\ ,
 \label{interaction}
\end{equation}
where $\vec{\sigma}_N$ is the nuclear spin operator of $^{207}$Pb, $\vec{\lambda}$ is the unit vector along the tetragonal axis of PbTiO$_3$ (axis $z$), $S$ is the Shiff moment of a nucleus under consideration, $X$ is determined by the electronic structure close to the nucleus:
\begin{equation}
    X=\frac{2\pi}{3}\left[\frac{\partial}{\partial z}\rho_{\psi}\left(\vec{r}\right)\right]_{x,y,z=0}\ ,
    \label{X}
\end{equation}
where  $\rho_{\psi}\left(\vec{r}\right)$  is an electronic density calculated from the relativistic wave function $\psi$ of a system under consideration. 

It follows from eq.~(\ref{interaction}) that for the interpretation of measurements in terms of NSM one should know $X$. Up to now only a few models for calculating $X$ were considered in crystals. In these models a local electronic structure in a vicinity of a heavy-atom nucleus was simulated by a cluster of nearest atoms and a system of point charges \cite{Ludlow:13} or by considering an effective state of Pb in oxygen environment \cite{Kuenzi:02, Mukhamedjanov:05}. 
In the present paper the $X$ parameter is evaluated using a new approach to study the core characteristics in solids. For comparison and accuracy estimation a value of $X$ for the $^1\Sigma^+$ state of a $^{207}$PbO molecule has also been considered using the same approximations made in solid state calculations and using two-component relativistic coupled clusters approach with single, double and perturbative triple cluster amplitudes (CCSD(T)).

\section{Method}

Solid state calculations can be efficiently performed using the Hartree-Fock method or density functional theory. In these approaches the wave function of a crystal is built as a Slater determinant of one-electron crystalline orbitals (COs) $\psi_i(\mathbf{r}, \mathbf{k})$ \footnote{Here the scalar-relativistic approximation is considered, i.e., the spin-orbit effects for explicitly treated electrons are usually excluded.}. The COs
   are formed as linear combinations
of Bloch functions $\varphi_\mu(\mathbf{r}, \mathbf{k})$. In the approximation of linear combination of atomic orbitals (LCAO), the Bloch functions
are written as (e.g., see \cite{Evarestov:07a}):
\begin{equation}
\varphi_\mu(\mathbf{r}, \mathbf{k})=\sum_{\mathbf{g}}\chi_\mu(\mathbf{r}-\mathbf{A_\mu}-\mathbf{g})e^{i\mathbf{k} \cdot \mathbf{g}}
\end{equation}
where $\mathbf{g}$ runs over all the lattice vectors,  $\mathbf{A_\mu}$ is the atomic coordinate in the zero reference cell on which $\chi_\mu$ is centered. By solving the Hartree-Fock or Kohn-Sham equations one obtains the CO-LCAO expansion coefficients $C_{\mu i}(\mathbf{k})$
for one-electron eigenstates $\psi_i(\mathbf{r}, \mathbf{k})$:

\begin{equation}
\psi_i(\mathbf{r}, \mathbf{k})=\sum_\mu C_{\mu i}(\mathbf{k})\varphi_\mu(\mathbf{r}, \mathbf{k})\ .
\end{equation}

%
%
One-electron reduced density matrix in a direct lattice takes the following form:

\begin{equation}
P^{\mathbf{g}-\mathbf{g\prime}}_{\mu\nu}=\sum_{\mathbf{k}} P_{\mu\nu}(\mathbf{k})e^{i\mathbf{k} \cdot (\mathbf{g}-\mathbf{g\prime})}\ ,
 \label{denmat}
\end{equation}
where $P_{\mu\nu}(\mathbf{k})$ is a density matrix in a reciprocal
  lattice and is determined by coefficients $C_{\mu i}(\mathbf{k})$.

To calculate the core characteristics in a heavy-element compound, a four-component relativistic approach is required in general. However, such four-component calculations are already complicated for molecular systems and become much more difficult for solids.

In the given study we have extended a two-step concept for calculation of heavy-atom core characteristics developed earlier by our group for molecules \cite{Titov:06amin} on the case of periodic structures (solids).
The new implementation of the concept is as follows.
Firstly, an electronic calculation for valence and outer-core electrons is performed using the DFT or Hartree-Fock method for a given crystal. Inner-core electrons of heavy atoms are excluded from calculations using the generalized relativistic effective core potential (GRECP) method \cite{Mosyagin:10a}. Secondly, since the inner-core parts of the valence one-electron ``pseudo-wavefunctions''
are smoothed in the GRECP calculations, they have to be recovered using some core-restoration method \cite{Titov:06amin}
before using them to evaluate the core characteristics.
The non-variational restoration, which is based on a proportionality of valence and virtual spinors in the inner-core region of heavy atoms, is used presently, in which one generates {\it equivalent} basis sets of one-center four-component spinors 
%
$$
  \left( \begin{array}{c} f_{nlj}(r)\theta_{ljm} \\
     g_{nlj}(r)\theta_{2j{-}l,jm} \\ \end{array} \right)
$$
%
%
 and smoothed two-component pseudospinors
$$
    \tilde f_{nlj}(r)\theta_{ljm}
$$
in all-electron Dirac-Fock(-Breit) and GRECP/SCF calculations of {the same} configurations of a considered atom and its ions \cite{HFDB, Bratzev:77, HFJ, Tupitsyn:95}.

In addition a basis set of one-component functions $\xi_p(\mathbf{x})$ is generated, where $\mathbf{x}$ denotes spatial and eigen-spin variables.  $\xi_p(\mathbf{x})$ are then expanded in the basis set of one-center two-component atomic {pseudospinors}
\begin{equation}
   \xi_p(\mathbf{x}) \approx
    \sum_{l=0}^{L_{max}}\sum_{j=|l-1/2|}^{j=|l+1/2|} \sum_{n,m}
    T_{nljm}^{p}\tilde f_{nlj}(r)\theta_{ljm}\ .
 \label{expansion}
\end{equation}

The atomic two-component pseudospinors are replaced by equivalent four-component spinors while the expansion coefficients from Eq.~(\ref{expansion}) are preserved:
\begin{equation} 
  \tilde{\xi}_p
  \approx
    \sum_{l=0}^{L_{\rm max}}\sum_{j=|l-1/2|}^{j=|l+1/2|} \sum_{n,m}
    T_{nljm}^{p}
     \left(
    \begin{array}{c}
    f_{nlj}(r)\theta_{ljm}\\
    g_{nlj}(r)\theta_{2j-l,jm}
    \end{array}
    \right)\ .
 \label{restoration}
\end{equation}

One-component $\chi_\mu(\mathbf{r}-\mathbf{A_\mu}-\mathbf{g})$ functions can be expanded in the basis of $\xi_p$ functions. This operation corresponds to a similarity transformation of the density matrix:
\begin{equation} 
  P^{\mathbf{g}}_{\mu\nu} \longrightarrow D^{\mathbf{g}}_{pq}=\tilde{D}^{\mathbf{g}}_{pq}
 \label{transf}
\end{equation}
where $D_{pq}$ ($\tilde{D}_{pq}$) is a density matrix in the basis of $\xi_p$ ($\tilde{\xi}_p$) functions.


A mean value of one-electron operator $A$ corresponding to a core property on a given atom in the zero reference cell can be evaluated as follows:
\begin{equation} 
   \langle {\bm{A}} \rangle\ =\ \sum_{\mathbf{g}} \sum_{pq} \tilde{D}^{\mathbf{g}}_{pq}
    {A}_{pq}\ ,
 \label{CoreProp} 
\end{equation} 
where ${A}_{pq}$ are matrix elements of operator $A$ in the basis of four-component functions 
$\tilde{\xi}_p$.
%

In the current implementation of restoration procedure the \textit{one-component} functions $\xi_p$  are taken in the form of contracted Gaussians that leads to analytical integration at step (\ref{transf}).

The developed code was interfaced to use the periodic density matrix (\ref{denmat}) calculated by {\sc crystal09} code \cite{CRYSTAL09b}.

To perform the GRECP/restoration evaluation of core characteristics in \textit{molecules} with the spin-orbit effects taken into account at the GRECP calculation stage,
  the code is also developed 
to the case when the \textit{two-component} molecular spinors $\varphi_\mu(\mathbf{x})$ are used.
The code is interfaced to use the two-component density matrices obtained in GRECP calculations using the {\sc dirac12} \cite{DIRAC12} and {\sc mrcc} \cite{Kallay:1,Kallay:3} codes.

\section{Results and discussions}

It was shown in a number of papers \cite{Cohen:06, Cohen:04, Bilc:08} that commonly used density functionals, such as PBE \cite{PBE} and PW91 \cite{PW91}, give extremely poor predictions of volume and strain of ferroelectrics such as PbTiO$_3$ and BaTiO$_3$.
A modified (WC) exchange density functional exploiting the generalized gradient approximation was proposed by Wu and Cohen in Ref.\ \cite{Cohen:06}. It was shown that the exchange WC functional significantly improves prediction of structural properties over the other popular functionals. More recently, a hybrid B1-WC functional was suggested
which gives rather good description of both structural and electronic properties of ferroelectric oxides \cite{Bilc:08}.
The latter functional was applied in the present paper for calculation of the valence electronic structure of tetragonal PbTiO$_3$, whereas the $1s-4f$ inner-core electrons of Pb were excluded from the calculation using the valence (semi-local) version of GRECP \cite{Mosyagin:10a} taken from our previous studies \cite{Petrov:13}.
To describe Pb in the PbTiO$_3$ crystal a new basis set consisting of $6s$, $7p$ and $4d$ segmentally-contracted Gauss functions was generated.
For Ti and O atoms the TZVP basis sets from Ref.~\cite{TZVPsolid} were used. Table \ref{PTOgeom} lists calculated bulk properties of a PbTiO$_3$ tetragonal phase in comparison with the experimental data \cite{Nelmes:85}.

\begin{table}[!h]
\caption{Equilibrium structural parameters for tetragonal\textit{ P4mm} structure of PbTiO$_3$. The positions of atoms \textit{u}$_z$ are given in terms of the lattice constants.}
\label{PTOgeom}
  \begin{tabular}{lll}
     \hline
     \hline
 Property & Calc. & Experiment 
\footnote[1]{Room temperature data, see Ref.~\cite{Nelmes:85}}   
 \\
 \textit{a}, $\AA$   & 3.84  & 3.90
   \\
 \textit{c / a}   & 1.12  & 1.07  \\
 \textit{u}$_z$(Pb),        & 0.000 &  0.000 \\   
 \textit{u}$_z$(Ti),        & 0.543 &  0.540
  \\
 \textit{u}$_z$(O1),        & 0.636 &  0.612 \\
 \textit{u}$_z$(O3),        & 0.137 &  0.112 \\

     \hline\hline
    \end{tabular}
    \end{table}


The computed value of effective T,P-odd interaction
of the $^{207}$Pb nucleus Schiff moment with the electronic density gradient, $\Delta\varepsilon$, as a function of the Pb shift with respect to O$_{12}$ cluster (with the simultaneous and proportional shift of Ti atoms) is shown in fig.~2.
\begin{figure}
\label{Xdist}
\includegraphics[scale=0.7]{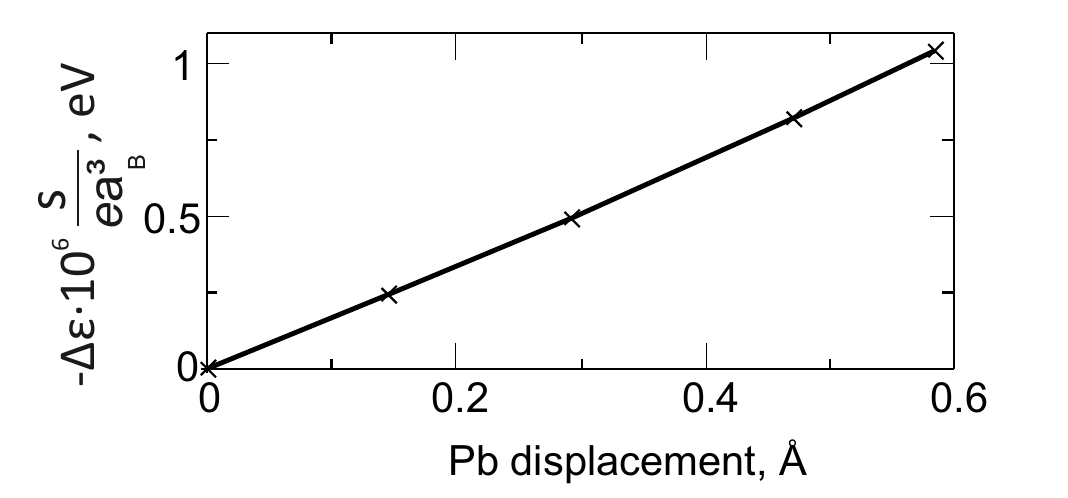}
\caption{$\Delta\varepsilon$ as a function of the Pb displacement.}
\end{figure}
At the experimental geometry we obtain  our final value, $\Delta\varepsilon=-0.82 \times 10^6\frac{S}{e a_B^3} eV$.

One should note that similar calculation with the Hartree-Fock method gives $\Delta\varepsilon=-1.34 \times 10^6\frac{S}{e a_B^3} eV$.
This is 2.3 times more than the value obtained in Hartree-Fock calculations using a cluster model of PbTiO$_3$ crystal in Ref.~\cite{Ludlow:13}. Interestingly that a more rough  estimation from Ref.~\cite{Mukhamedjanov:05} ($-1.1 \times 10^6\frac{S}{e a_B^3} eV.$) is closer to our final value. It is shown below that the Hartree-Fock approximation significantly overestimates also the $X$ value in a molecular PbO case.

As it follows from previous considerations of PbTiO$_3$ \cite{Cohen:04,Cohen:06,Bilc:08} and other systems there are no reliable \textit{theoretical} criteria to choose the most appropriate DFT exchange-correlation functional, therefore, the error of the evaluated core characteristics can hardly be estimated on the basis of only the DFT studies.
Another source of possible uncertainties of the present solid state calculations is the neglect of spin-orbit effects for the valence and explicitly treated outer-core electrons of Pb at the GRECP calculation stage with the exploited non-relativistic codes for periodic structures (though, they are partly recovered at the restoration stage).
The theoretical estimation of errors for used DFT approximations can be obtained by comparing them to the results of high-level correlation calculations with explicit inclusion of spin-orbit effects. In our previous studies the $X$ parameter was evaluated for TlF \cite{Petrov:02} and RaO \cite{Kudashov:13} molecules using the above mentioned two-step procedure and relativistic Fock-space coupled cluster method with single and double cluster amplitudes.
The influence of the inner core $-$ valence electron correlations on the $X$ value was estimated in \cite{Dzuba:02} to be no more than 2\% for TlF. In the present paper we have calculated the $X$ parameter for the PbO molecule in its $^1\Sigma^+$ ground state at different levels of correlation treatment. 
%
%
Two series of calculations were performed: (i) without accounting for the spin-orbit term in the GRECP operator (see \cite{Mosyagin:10a}) and (ii) with this term. The former corresponds to a scalar-relativistic approximation (one-component) for valence and outer core electrons 
\footnote{It should be stressed that the most important part of the spin-orbit effects is taken into account for all the electrons at the restoration stage even in the ``scalar-relativistic'' approximation.}%
, the latter includes the spin-orbit term and assumes the two-component description of one-electron functions. The results of calculations are listed in Table~\ref{TDFTResults}.

\begin{table}[!h]
\caption{
The values of $X$(PbO) calculated using popular exchange-correlation functionals in comparison with high-level coupled clusters calculations. The GRECP calculations were performed with (1c) and without (2c) taking into account the spin-orbit effects.
}
\label{TDFTResults}
\begin{tabular}{ l  c  c }
\hline\hline
Method & $X (1c)$     & $X (2c)$ \\
\hline
  Hartree-Fock    & {} 9324 & {} 9436  \\  
\hline
  LDA    & {} 6688 & {} 7144  \\  
  PBE \cite{Perdew:96} & {} 6735  & {} 7184  \\
  B3LYP \cite{b3lyp} &  {} 7316 & {} 7725  \\
  PBE0 \cite{pbe0}   & {} 7599 & {} 8020  \\
  WC  \cite{Cohen:06}   & {} 6829 & {}   \\   
  B1-WC \cite{Bilc:08}   & {} 7400  & {}   \\   
\hline  
  CCSD    & {} 7699 & {}  8076 \\    
  CCSD(T)    & {} 7489 & {} 7875  \\    
\hline\hline
\end{tabular}
\end{table}


It follows from table \ref{TDFTResults} that the spin-orbit contribution to $X$ at the CCSD(T) level is about 5\%. The scalar-relativistic B1-WC calculation of $X$ parameter reproduces the scalar-relativistic CCSD(T)
  value
almost exactly. Moreover, all of the density functional based descriptions also reproduce the spin-orbit contributions with good accuracy. In contrast, the Hartree-Fock method gives rather poor prediction of both spin-orbit contribution and the total value.
A correlation contribution estimated as a difference between the DFT and HF results is about three times bigger for the solid state case   (see above).
This can indirectly explain the failure of describing the properties of PbTiO$_3$ ferroelectric phase by the most popular exchange-correlation functionals mentioned in \cite{Cohen:06}.
Taking into account these results and a more complicated structure of PbTiO$_3$ crystal one can expect an error of 15\% for the solid-state results.


\section{Conclusion}

A new method that is based on the two-step concept of calculation of core-localized characteristics in solids is proposed and implemented.
%
%
The developed approach
is applied to evaluation of the $X$ parameter of T,P-odd interaction in a periodic model of the PbTiO$_3$ crystal. The calculated interaction energy is found to be $-0.82 \times 10^6\frac{S}{e a_B^3} eV$. The accuracy of the developed method for the case of DFT treatment of electronic correlations is investigated
and estimated as 15\% for the considered system. It can be increased further by explicit treatment of correlation effects in solids within the coupled clusters approach, though this method is not generally available to-date for calculation of periodic systems.

The approach can be used to calculate a number of other core properties and parameters such as hyperfine structure constants, chemical shifts, etc.\ in solids using well-developed nonrelativistic packages for calculation of periodic structures such as {\sc crystal09} \cite{CRYSTAL09b}.


\section*{Acknowledgement}
This work is supported by the SPbU Fundamental Science Research grant from Federal budget No.~0.38.652.2013 and RFBR Grant No.~13-02-01406. L.S.\ is also grateful to the Dmitry Zimin ``Dynasty'' Foundation.  The {\sc crystal09} calculations were performed at the Supercomputer ``Lomonosov''.


\end{document}